\documentclass[twocolumn,english]{revtex4-1}

\usepackage[T1]{fontenc}
\usepackage{amsbsy}
\usepackage{amsmath}
\usepackage{amstext}
\usepackage{amssymb}
\usepackage{graphicx}
\usepackage{units}
\usepackage{babel}

\begin{document}
\title{Population inversion in Landau-quantized graphene}
\author{Florian Wendler}
\email{florian.wendler@tu-berlin.de}
\author{Ermin Malic}
\affiliation{Institute of Theoretical Physics,
Nonlinear Optics and Quantum Electronics, Technical University Berlin,
Hardenbergstrasse 36, Berlin 10623, Germany.}


\begin{abstract}
Landau level lasers have the advantage of tunability of the laser frequency by
means of the external magnetic field. The crucial prerequisite of such a laser
is a population inversion between optically coupled Landau levels. Efficient
carrier-carrier and carrier-phonon scattering generally suppresses this effect
in conventional materials. Based on microscopic calculations, we predict for the
first time the occurrence of a long-lived population inversion in
Landau-quantized graphene and reveal the underlying many-particle mechanisms. To
guide the experimental demonstration, we present optimal conditions for the
observation of a maximal population inversion in terms of experimentally
accessible parameters, such as the strength of the magnetic field, pump fluence,
temperature, and doping. We reveal that in addition to the tunability of the
Landau-level laser frequency, also the polarization of the emitted light can be
tuned via gate voltage controlling the doping of the sample.
\end{abstract}

\maketitle

In 1986 H. Aoki proposed the first Landau-level
laser for two-dimensional (2D) electron systems \cite{Aoki1986} exploiting
the discreteness of the Landau levels (LLs) to tune the laser frequency
through the magnetic field. The key challenge for the realization
of such a LL laser is to obtain and to sustain a long-lived population
inversion (PI) between LLs. This is difficult to achieve in conventional
semiconductors where strong Coulomb scattering between equidistant
LLs acts in favor of an equilibrium Fermi-Dirac distribution. Similarly,
phonon-induced scattering can counteract PI, if the phonon energy
is in resonance with the inter-Landau level transitions involved in
the lasing process.

\begin{figure}[t!]
\begin{centering}
\includegraphics[width=0.95\linewidth]{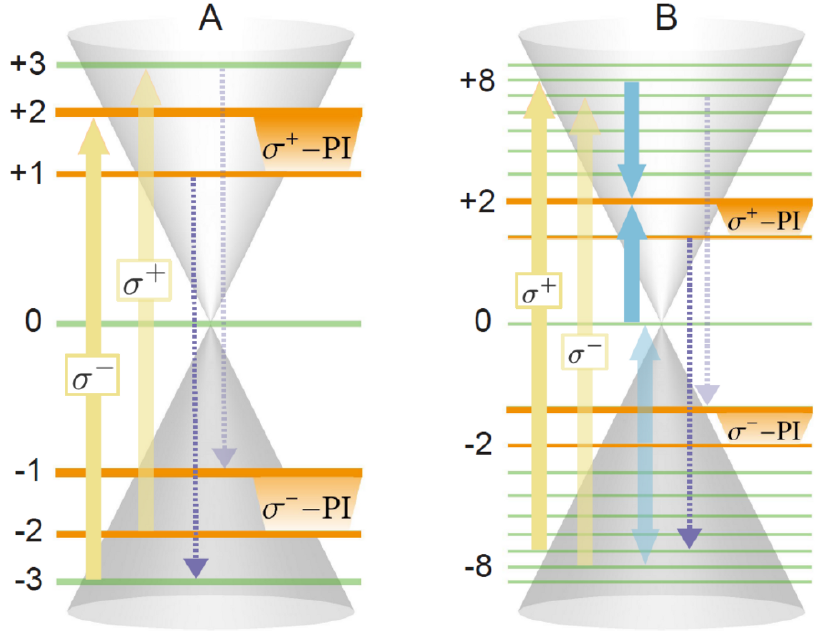}
\par\end{centering}

\caption{\textbf{Two routes to population inversion (PI) in Landau-quantized
graphene. }The sketches show graphene's energetically lowest Landau
levels (LLs) with the Dirac cone in the background. They illustrate
different schemes to achieve population inversion between $\text{LL}_{+1}$
and $\text{LL}_{+2}$ ($\sigma^{+}$-PI) and between $\text{LL}_{-2}$
and $\text{LL}_{-1}$ ($\sigma^{-}$-PI), respectively. Scheme A is
based solely on optical pumping (indicated by the yellow arrows),
which creates PI by populating $\text{LL}_{+2}$ and depopulating
$\text{LL}_{-2}$. Scheme B exploits efficient Auger scattering (cf.
blue arrows) between the equidistant LLs $+8,+2,$ and $0$ to redistribute
optically pumped electrons (holes) from $\text{LL}_{+8}$ ($\text{LL}_{-8}$)
to $\text{LL}_{+2}$ ($\text{LL}_{-2}$). While the mechanism A is
the basis for a three-LL laser system, the mechanism B involves a
fourth level, which might be of advantage for the efficiency of such
a laser. To allow for continuous laser action, phonon-assisted relaxation
channels are needed that connect all LLs involved in the laser system
(cf. purple-dotted arrows). The processes building up the $\sigma^{+}$-PI
in the conduction band are illustrated by bold arrows, while those
giving rise to the $\sigma^{-}$-PI in the valence band are transparent.}
\label{fig:gain-sketch}
\end{figure}
Graphene as a two-dimensional zero-gap semiconductor with remarkable
properties \cite{Novoselov2004,Zhang2005,Geim2007} offers optimal
conditions for LL lasing. Its linear electronic dispersion leads to
an unconventional non-equidistant LL spacing including the appearance
of a zero Landau level in an external magnetic field \cite{Sadowski2006,Plochocka2008,Orlita2008}.
The observation of a number of interesting effects such as the fractional
quantum Hall effect \cite{Du2009,Bolotin2009}, a giant Faraday rotation
\cite{Crassee2011}, the quantum ratchet effect \cite{Drexler2013},
the Hofstadter butterfly \cite{Ponomarenko2013,Dean2013,Hunt2013},
and the demonstration of a tunable THz detector \cite{Kawano2013}
has already attracted enormous interest to Landau-quantized graphene
\cite{Goerbig2011Review}. The non-equidistant LLs and the specific
optical selection rules allowing transitions between LLs with $n\rightarrow n\pm1$
\cite{Sipe2012} make graphene an optimal material for the realization
of an efficient two-dimensional LL laser. A transient population inversion
in graphene without a magnetic field has already been theoretically
predicted \cite{Ryzhii2007,Winzer2013gain} and experimentally demonstrated
\cite{Li2012,Boubanga-Tombet2012,Gierz2013}. It emerges as a result
of a relaxation bottleneck close to the Dirac point and decays mainly
due to Coulomb-induced recombination processes \cite{Winzer2013gain}.

In this article, we predict the occurrence of a long-lived population
inversion in Landau-quantized graphene. We present two different experimentally
feasible mechanisms to achieve the population inversion induced by
optical pumping, cf. Fig. \ref{fig:gain-sketch}. The first mechanism
(A) is based on the specific optical selection rules in Landau-quantized
graphene yielding the possibility to selectively pump a single LL
transition constituting an effectively three-level laser system, cf.
Fig. \ref{fig:gain-sketch}A. The second mechanism (B) exploits the
scattering among electrons to achieve PI and thereby adds an additional
level to the system, which can be beneficial for the efficiency of
the laser, cf. Fig. \ref{fig:gain-sketch}B. Interestingly, scheme
B provides a Coulomb-induced mechanism to create PI, which is quite
remarkable, because Coulomb-induced Auger scattering was shown to
rather reduce PI in graphene \cite{Winzer2013gain} and has been believed
to be the main obstacle for the realization of a graphene-based two-dimensional
LL laser \cite{Plochocka2009}. Since we preserve the electron-hole
symmetry, PI is obtained in the conduction band and in the valence
band at the same time. It occurs between the LLs with the indices
$n=1$ and $n=2$. However, we want to stress that also other schemes
are possible to obtain PI for different LL transitions. Note that
the PI transitions in the conduction and in the valence band are optically
coupled by inversely circularly polarized photons, i.e. photons created
in a stimulated emission process inducing the electronic transition
$\text{LL}_{+2}\rightarrow\text{LL}_{+1}$ ($\text{LL}_{-1}\rightarrow\text{LL}_{-2}$)
are $\sigma^{+}$- polarized ($\sigma^{-}$- polarized). Hence, we
label the corresponding population inversion as $\sigma^{+}$- PI
and $\sigma^{-}$- PI, respectively. The proposed PI schemes A and
B produce circular polarized photons with both rotational directions,
which can be combined to produce linear polarized laser light.

\begin{figure}[t]
\begin{centering}
\includegraphics[width=0.95\linewidth]{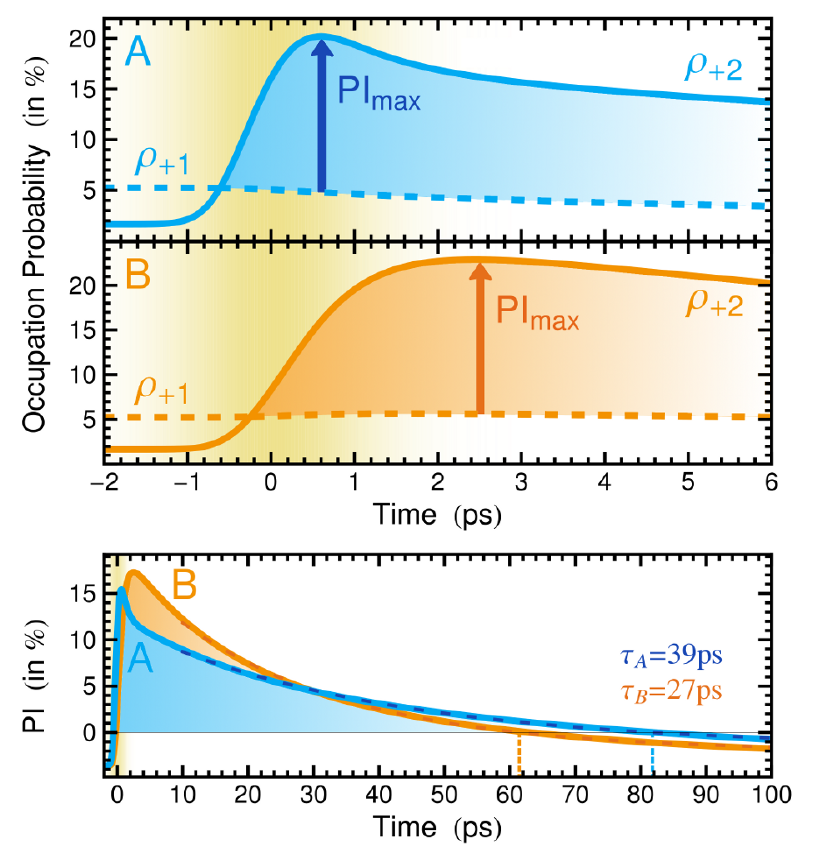}
\par\end{centering}

\caption{\textbf{LL occupations involved in the build-up of the population
inversion.} The temporal evolution of the occupations $\rho_{+1}$
and $\rho_{+2}$ are shown in the upper panel for the PI schemes A
and B (depicted in Fig. \ref{fig:gain-sketch}) illustrating the occurrence
of a pronounced population inversion defined by
$\text{PI}=\rho_{+2}-\rho_{+1}>0$
(blue and orange shaded areas). The system is assumed to be at room
temperature and under a magnetic field of $B=\unit[4]{T}$. It is
optically excited  using a pulse  with a width of $\unit[1]{ps}$
(cf. yellow area in the background) and a fluence of
$\epsilon_{\text{pf}}=\unit[1]{\mu Jcm^{-2}}$
in scheme A and $\epsilon_{\text{pf}}=\unit[2.27]{\mu Jcm^{-2}}$
in scheme B to ensure the same pulse area. The lower panel shows the
temporal evolution of the PI featuring an ultrafast build-up and a
slow decay on a ps time-scale. Exponential fitting (dashed lines)
reveals the corresponding decay times of $\tau_{A}=\unit[39]{ps}$
and $\tau_{B}=\unit[27]{ps}$ for both PI schemes. The vertical dashed
lines mark the points in time where the PI vanishes ($\text{PI}=0$)
in the respective scheme.}
\label{fig:occupations}
\end{figure}
With respect to future graphene-based applications, microscopic insights
into the carrier dynamics in graphene are of crucial importance. While
the relaxation channels without a magnetic field have been already
thoroughly studied in experiment \cite{Dawlaty2008,SunNorris2008,Winnerl2011,Breusing2011,Johannsen2013,Gierz2013,Brida2013}
and theory \cite{Rana2007,Winzer2010_Multiplication,Winzer2013gain,MalicBuch,Brida2013,Kadi2014},
the investigation of the carrier dynamics in Landau-quantized graphene
has just started to pick up pace very recently \cite{Plochocka2009,Li2013,Wang2013,Wendler_Phonon_2013,Wendler_CM_2014,DTS_MittendorfWendler2013}.
We have developed a theory based on the density matrix approach \cite{KochBuch,MalicBuch}
providing access to time and energy-dependent relaxation dynamics
in Landau-quantized graphene and revealing microscopic insights into
the underlying many-particle scattering pathways \cite{Wendler_Phonon_2013,Wendler_CM_2014}.
The temporal evolution of LL carrier occupations $\rho_{i}=\left\langle a_{i}^{+}a_{i}\right\rangle $
and microscopic polarizations $p_{ij}=\left\langle a_{i}^{+}a_{j}\right\rangle $
(with the fermionic creation and annihilation operators $a_{i}^{+}$
and $a_{i}$) determining the strength of optical LL transitions is
obtained using the graphene Bloch equations in the presence of a magnetic
field 
\begin{widetext}
\begin{eqnarray}
\dot{\rho}_{i}(t) & = & -2\sum_{j}\text{Re}[\Omega_{ij}(t)\,
p_{ij}(t)]+S_{i}^{\text{in}}(t)\left[1-\rho_{i}(t)\right]-S_{i}^{\text{out}}
(t)\rho_{i}
(t) ,\label{eq:Bloch-1}\\
\dot{p}_{ij}(t) & = &
\left[i\triangle\omega_{ij}-\gamma(t)\right]p_{ij}(t)+\Omega_{ij}(t)\,\left[
\rho_{i}(t)-\rho_{j}(t)\right].\label{eq:Bloch-2}
\end{eqnarray}
\end{widetext}
The set of coupled differential equations has been obtained by exploiting
a correlation expansion within the second-order Born-Markov approximation
\cite{KochBuch,MalicBuch}. Here, we explicitly take into account
the occupations and polarizations of the energetically lowest LLs
up to $n=10$, including the optical excitation as well as all energy-conserving
carrier-carrier and carrier-phonon scattering processes. The introduced
indices  $i=(\xi_{i},\lambda_{i},n_{i},m_{i})$ are compound indices
comprising the valley $\xi_{i}$, the band $\lambda_{i}$, the Landau
level index $n_{i}$, and the quantum number $m_{i}$ being connected
to the centers of the cyclotron orbits in the graphene plane \cite{Goerbig2011Review}.
The Rabi frequency $\Omega_{ij}=\frac{e_{0}}{m_{0}}\boldsymbol{M}_{ij}\cdot\boldsymbol{A}(t)$
appearing in Eqs. \ref{eq:Bloch-1} and \ref{eq:Bloch-2} depends
on the electron's charge $e_{0}$, its free mass $m_{0}$, the optical
matrix element $\boldsymbol{M}_{ij}$ \cite{Sipe2012}, and on the
vector potential $\boldsymbol{A}(t)$. The explicitly time-dependent
scattering rates $S_{i}^{\text{in/out}}(t)$ incorporate all energy-conserving
electron-electron and electron-phonon scattering processes including
time-dependent Pauli blocking terms. The Coulomb interaction is dynamically
screened taking into account the momentum dependence of the dielectric
function in the random phase approximation \cite{Wendler_CM_2014,Goerbig2011Review}.
Phonon-induced scattering via the dominant optical phonon modes $\Gamma\text{TO}$,
$\Gamma\text{LO}$, and $\text{KTO}$ is taken into account, where
a coupling to a bath is considered. The microscopic polarization decays
due to a dephasing $\gamma(t)$ caused by many-particle scattering
as well as impurity-induced LL broadening. The energy difference
$\triangle\omega_{ij}=(\epsilon_{i}-\epsilon_{j})/\hbar$
between $\text{LL}_{i}$ and $\text{LL}_{j}$ describes the oscillation
of the corresponding polarization. Excitonic effects are not considered,
since they are known to be weak in the low-energy limit near the Dirac
points of graphene \cite{Chae2011,Mak2011}, which appears to be also
valid under Landau-quantization, where no signatures of excitonic
effects were observed in the low-energy regime \cite{Sadowski2006,Orlita2008}.
Furthermore, we assume an impurity-induced broadening of the LLs calculated
in a self-consistent Born approximation \cite{Ando1974_I,Wendler_CM_2014},
where a reasonable strength of the impurity scattering is chosen \cite{Ando1998,Levitov2012,DTS_MittendorfWendler2013}
yielding a broadening of approximately $\unit[4]{meV}$.

\begin{figure}[t]
\begin{centering}
\includegraphics[width=0.95\linewidth]{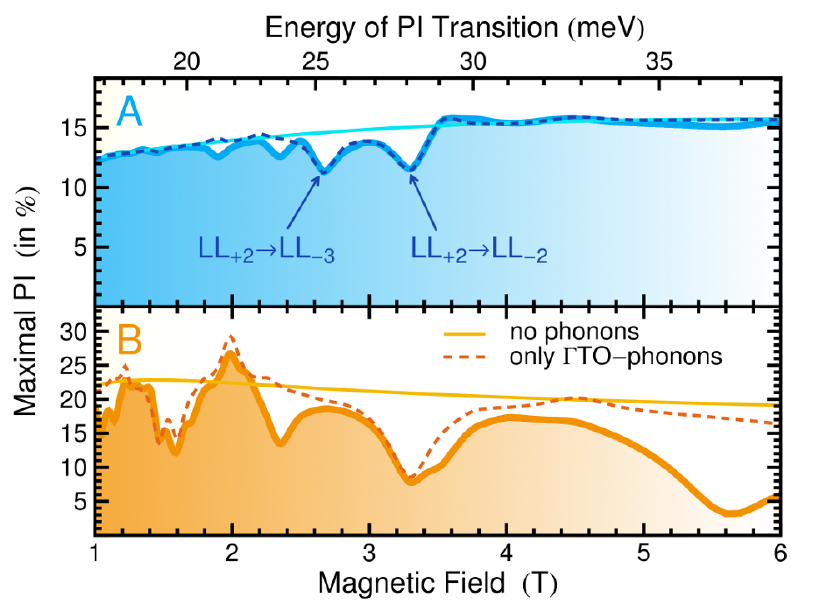}
\par\end{centering}

\caption{\textbf{Maximal} \textbf{population inversion as a function of magnetic
field.} The dependence on the magnetic field is characterized by pronounced
peaks and dips at resonant conditions, where an optical phonon energy
matches the spacing between two Landau levels involved in the respective
PI scheme. To illustrate this, the thin lines in A and B show the
maximal PI, when electron-phonon scattering is switched off, and the
dashed lines are obtained considering only $\Gamma\text{TO}$-phonons.
In scheme A, two dips are present at $B=\unit[2.67]{T}$ and $B=\unit[3.30]{T}$
corresponding to resonances of the $\Gamma\text{TO}$-phonon energy
with the transitions $\text{LL}_{+2}\rightarrow\text{LL}_{-3}$ and
$\text{LL}_{+2}\rightarrow\text{LL}_{-2}$, respectively. In scheme
B, three dips and two peaks appear and can be unambiguously ascribed
to specific phonon-assisted LL transitions (cf. the text). In general,
providing additional decay channels the carrier-phonon scattering
reduces the PI. However, relaxation channels can also increase the
PI, if they lead to a decrease of $\rho_{+1}$ (lower PI level), an
increase of $\rho_{+2}$ (upper PI level), or if they populate $\text{LL}_{-7}$
that increases the efficiency of pumping.}
\label{fig:B-dependence}
\end{figure}

We investigate the carrier dynamics in graphene in the presence of
an external magnetic field of $B=\unit[4]{T}$. We consider the system
to be at room temperature with initial Fermi-Dirac distributed occupations
that are optically excited by a pump pulse with a width of $\unit[1]{ps}$,
a pump fluence of $\epsilon_{\text{pf}}=\unit[1]{\mu Jcm^{-2}}$,
and an energy matching the pumped LL transitions of the respective
PI scheme, cf. Fig. \ref{fig:gain-sketch}. In order to keep the excitation
energy-dependent pulse area constant, the pump fluence in scheme B
is increased to $\epsilon_{\text{pf}}=\unit[2.27]{\mu Jcm^{-2}}$.
Solving the graphene Bloch equations (Eqs. \ref{eq:Bloch-1} and \ref{eq:Bloch-2})
yields the temporal evolution of the microscopic polarizations $p_{ij}$
and of the LL occupations $\rho_{i}$ allowing us to investigate the
interplay between optical transitions on the one side and carrier-carrier
and carrier-phonon scattering processes on the other side. Since neutral
Landau-quantized graphene is symmetric for electrons and holes, we
focus the discussion on the $\sigma^{+}$-PI in the conduction band.
Figure \ref{fig:occupations} illustrates the time-dependent occupations
of $\text{LL}_{+1}$ and $\text{LL}_{+2}$ (upper panel) as well as
the resulting population inversion (lower panel). The qualitative
behavior is the same in both PI schemes: While $\rho_{+1}(t)$ changes
only slightly, $\rho_{+2}(t)$ shows a fast increase on a sub-picosecond
time scale during and shortly after the optical excitation (illustrated
by the yellow area in the background) followed by a slow decay on
a picosecond time scale. We find a long-lived population inversion
that is defined by 
\begin{equation}
\text{PI}=\rho_{+2}-\rho_{+1}>0\label{eq:PI}
\end{equation}
and represented by the respective areas between $\rho_{+1}$ and $\rho_{+2}$
in Fig. \ref{fig:occupations}. Exponential fits to the temporal evolutions
of PI for schemes A and B (dashed lines in lower panel of Fig. \ref{fig:occupations})
reveal their decay times $\tau_{A}=\unit[39]{ps}$ and $\tau_{B}=\unit[27]{ps}$.
Since scheme A provides a straightforward approach to induce PI as
a direct consequence of the optical excitation, the maximal value
$\text{PI}_{\text{max}}^{\text{A}}=15.4\%$ is reached already during
pumping. In scheme B, on the other hand, Coulomb-scattering is needed
to induce PI by redistributing the optically excited charge carriers,
consequently, the build-up time is longer and its maximum $\text{PI}_{\text{max}}^{\text{B}}=17.3\%$
is reached with a delay of a few picoseconds. While both PI schemes
are suited to create a significant PI with a rather long decay time,
the advantage of scheme B is its additional fourth level making it
a potentially better laser system. This is reflected by its higher
maximal PI value resulting from the fact that the optical excitation
does not directly induce an enhanced occupation of the upper PI level
$\text{LL}_{+2}$, but an additional level $\text{LL}_{+8}$ is used
(cf. Fig. \ref{fig:gain-sketch}). As a consequence, optically excited
charge carriers in $\text{LL}_{+8}$ scatter down to $\text{LL}_{+2}$
already during pumping which reduces the pumping saturation and allows
the excitation of more charge carriers. The cost of the enhanced maximal
PI is a faster decay: In scheme A the PI vanishes after $\sim\unit[82]{ps}$,
while in scheme B the PI lasts $\sim\unit[62]{ps}$. This can be explained
by the more complex PI scheme B involving more LLs and therefore opening
up more possible phonon-assisted decay channels reducing the PI (cf.
Fig. \ref{fig:B-dependence}).

To obtain more insights into the underlying elementary processes and
to guide future experiments towards the demonstration of LL lasers,
we investigate in the following the maximal PI as a function of experimentally
accessible quantities, such as magnetic field strength, pump fluence,
temperature, and doping. Figure \ref{fig:B-dependence} shows the
dependence on the magnetic field for both PI schemes, where the thick
lines represent the full dynamics, while the thin lines describe the
dynamics without phonon-assisted scattering, and the dotted-thin lines
taking into account only the impact of $\Gamma\text{TO}$-phonons.
For this investigation, we determine the maximal PI at different magnetic
fields $B$, while the pulse area is held constant by scaling the
pump fluence linearly with the magnetic field. The energy of
the PI transition $\epsilon_{+2}-\epsilon_{+1}$ is tunable via the
magnetic field and is given on the upper axis. The dynamics without
phonons (thin lines) shows a weak dependence on the magnetic field:
At higher $B$, the LLs are shifted to higher energies, thereby reducing
the initial occupations in the conduction band, in particular $\rho_{+1}$
decreases resulting in a stronger PI. However, at the same time the
scattering generally becomes more efficient with increasing magnetic
fields, since the degeneracy of LLs scales with $B$. This enhances
the dephasing $\gamma(t)$ of $p_{ij}$ (cf. Eq. \ref{eq:Bloch-2})
and reduces the pumping efficiency and consequently also the maximal
PI. These counteracting effects nearly balance each other out resulting
in a very weak dependence on the magnetic field. Switching on phonon-induced
scattering, the dependence qualitatively changes and pronounced peaks
and dips emerge in the $B$-dependence of the maximal PI, cf. thick
lines in \ref{fig:B-dependence}. They indicate magnetic fields that
fulfill the resonance condition between the energy of an optical phonon
and inter-Landau level transitions involved in the respective PI scheme.
This is further evidenced by showing the dynamics that only includes
$\Gamma\text{TO}$-phonons (dashed lines), where the number of peaks
and dips is clearly reduced. In scheme A, two main resonances are
present at the magnetic fields $B=\unit[2.67]{T}$ and $B=\unit[3.30]{T}$.
The first corresponds to the transition $\text{LL}_{+2}\rightarrow\text{LL}_{-3}$,
while the second coincides with the transition $\text{LL}_{+2}\rightarrow\text{LL}_{-2}$,
both reducing the PI by providing direct decay channels of the PI.
The $B$-dependence in scheme B is more complex, since more LLs are
involved (cf. Fig. \ref{fig:gain-sketch}) and hence more resonances
occur. The three distinct dips at $B=\unit[1.45]{T}$, $B=\unit[1.60]{T}$,
and $B=\unit[3.30]{T}$ correspond to the transitions $\text{LL}_{+2}\rightarrow\text{LL}_{-8}$,
$\text{LL}_{+2}\rightarrow\text{LL}_{-7}$, and $\text{LL}_{+2}\rightarrow\text{LL}_{-2}$
(and at the same time $\text{LL}_{+8}\rightarrow\text{LL}_{0}$),
respectively. For reasons of clarity, symmetric hole transitions are
omitted for the discussion. Besides the decay of the PI similar to
scheme A, phonons further reduce the PI in scheme B as they compete
with the Coulomb channels, which are responsible for the appearance
of the PI.\textbf{ }Interestingly, phonon-induced scattering can also
increase the PI, as can be seen at $B=\unit[1.22]{T}$ and $B=\unit[1.99]{T}$.
At these magnetic field strengths, the energy of the $\Gamma\text{TO}$-phonon
is in resonance with the transitions $\text{LL}_{+4}\rightarrow\text{LL}_{-7}$
and $\text{LL}_{+1}\rightarrow\text{LL}_{-7}$, respectively. The
former case positively affects the pumping, while in the latter case
the PI is not only increased through the depletion of $\text{LL}_{+1}$.
It also couples the lower laser level $\text{LL}_{+1}$ with the ground
state $\text{LL}_{-7}$ from which electrons are excited and thus
connects the states of the four-level laser system (cf. Fig. \ref{fig:gain-sketch}).
This opens up the possibility of continuous laser action in this system. 

\begin{figure}[t]
\begin{centering}
\includegraphics[width=0.95\linewidth]{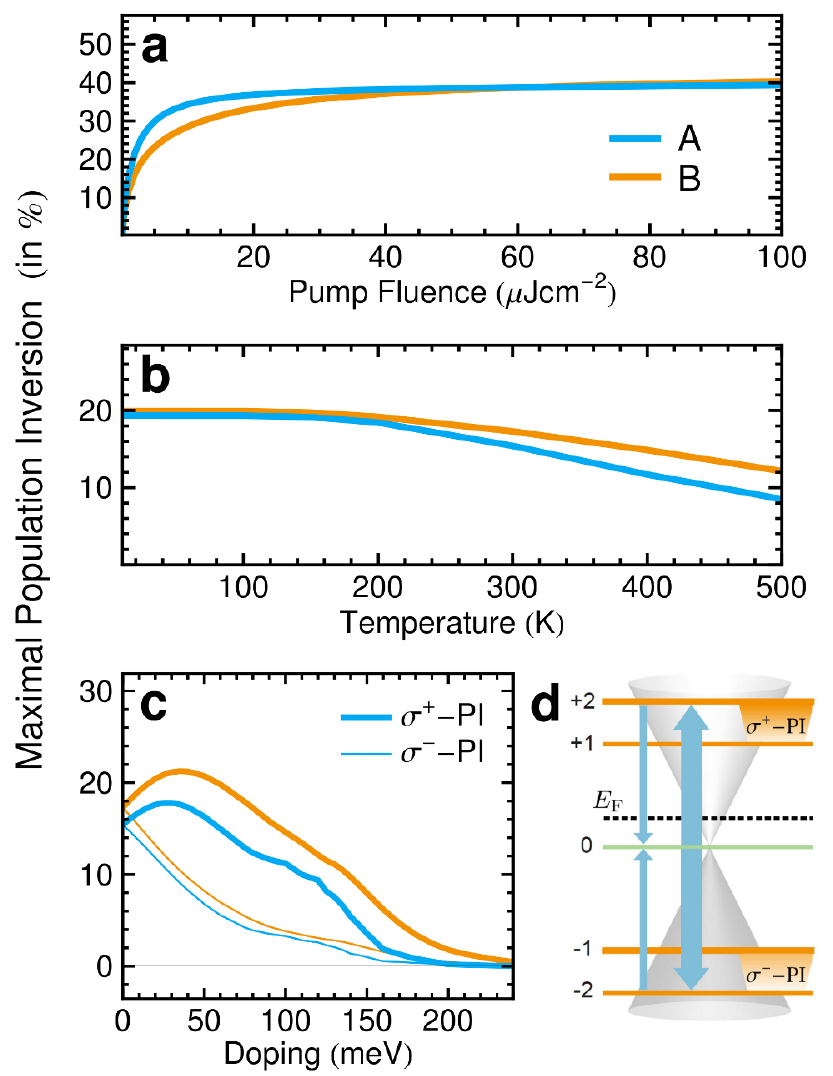}
\par\end{centering}

\caption{\textbf{Maximal population inversion in dependence of pump fluence,
temperature, and doping.} The maximal PI between $\text{LL}_{+1}$
and $\text{LL}_{+2}$ is plotted as a function of (a) pump fluence,
(b) temperature, and (c) doping. In the latter case, we also show
the PI between $\text{LL}_{-2}$ and $\text{LL}_{-1}$, since the
electron-hole symmetry is broken once a doping is introduced. The
sketch (d) illustrates the Coulomb-induced scattering (thick blue
arrows) prevailing against its inverse process (thin blue arrows)
that result in the asymmetry of the two PI transitions.}
\label{fig:dependencies}
\end{figure}

Besides the strength of the magnetic field, the pump fluence, the
temperature, and the doping can be controlled in the experiment. To
reveal the optimal conditions for a maximal population inversion,
we perform further calculations investigating the impact of these
externally accessible parameters, cf.  Fig. \ref{fig:dependencies}.
Since the optical excitation generates non-equilibrium carriers that
fill up the upper gain level $\text{LL}_{+2}$, the PI increases with
the pump fluence, cf. Fig. \ref{fig:dependencies}a\textbf{.} A saturation
is reached when the optically excited LL becomes half filled, which
hinders further pumping and imposes an upper limit to the PI. The
initial thermal occupation of $\rho_{+1}$ further decreases the maximal
PI resulting in a saturation value of approximately $40\%$ in scheme
A and $45\%$ in scheme B (not shown in the figure). The maximal PI
at low fluences is slightly larger in scheme A, since the pulse area
is larger in scheme A due to the lower frequency of the pump pulse.
The higher saturation value of the PI in scheme B is a manifestation
of the advantage of a four-level laser system compared to a three-level
system in scheme A. 

Furthermore, the PI is inversely correlated to the temperature $T$,
i.e. the higher $T$ the smaller is the population inversion, cf.
Fig. \ref{fig:dependencies}b. At higher temperatures, the initial
carrier occupation $\rho_{+1}(t_{0})$ increases faster than $\rho_{+2}(t_{0})$
resulting in a less pronounced PI, cf. Eq. \ref{eq:PI}. For temperatures
less than $\unit[160]{K}$ the initial difference $\rho_{+1}(t_{0})-\rho_{+2}(t_{0})$
is below $0.5\%$, however it increases to $3.6\%$ at room temperature
and reaches $7.1\%$ at $T=\unit[500]{K}$ which exactly reflects
the predicted temperature dependence of the maximal PI. In scheme
A, the effect is more pronounced, because the temperature also affects
the LL occupations $\rho_{+2}(t_{0})$ and $\rho_{+3}(t_{0})$ that
are involved in the optical excitation, whereas this effect is negligible
in scheme B with the energetically high $\text{LL}_{+8}$ and $\text{LL}_{-7}$. 

Finally, doping opens up the possibility to control the PI by shifting
the Fermi energy away from $E_{\text{F}}=0$, which can be achieved
by applying a gate voltage. This breaks the electron-hole symmetry
and allows to tune the relative population inversion between the two
inversely polarized LL transitions $\text{LL}_{\pm2}\rightarrow\text{LL}_{\pm1}$,
cf. Fig. \ref{fig:dependencies}c. While a small positive Fermi energy
gives rise to an increase of the $\sigma^{+}$- PI, the impact on
the $\sigma^{-}$- PI is opposite. This behavior can be attributed
to new scattering channels that are forbidden under electron-hole
symmetry, but arise as soon as this symmetry is broken. For simplicity,
we focus on the simple PI scheme A in the following: An up-shift of
the Fermi energy results in a less efficient pumping of the transition
$\text{LL}_{-3}\rightarrow\text{LL}_{+2}$ due to an enhanced Pauli
blocking in comparison to the transition $\text{LL}_{-2}\rightarrow\text{LL}_{+3}$.
According to this, $\sigma^{+}$- PI should be suppressed, while $\sigma^{-}$-
PI is expected to be enhanced. Interestingly, we observe the opposite
behavior, as shown in Fig. \ref{fig:dependencies}. To understand
the doping dependence, we consider the energy-conserving Coulomb process
involving the transitions $\text{LL}_{0}\rightarrow\text{LL}_{+2}$
and $\text{LL}_{0}\rightarrow\text{LL}_{-2}$ (outward scattering),
which is canceled out in an electron-hole symmetric system that also
exhibits the inverse process (inward scattering: $\text{LL}_{+2}\rightarrow\text{LL}_{0}$
and $\text{LL}_{-2}\rightarrow\text{LL}_{0}$) occurring with the
same probability. However, due to the asymmetric pumping and a more
than half-filled $\text{LL}_{0}$ in a n-doped sample, the Coulomb-induced
outward scattering prevails over the inward scattering, cf. Fig. \ref{fig:dependencies}d.
As a result, $\sigma^{+}$- PI is enhanced, while $\sigma^{-}$- PI
is suppressed, as observed in Fig. \ref{fig:dependencies}c. Shifting
the Fermi energy further away from the neutral position, the PI of
both transitions decreases, which can be readily understood considering
the initial occupations. When the Fermi energy reaches the vicinity
of $\text{LL}_{+1}$, its initial occupation $\rho_{+1}(t_{0})$ is
considerably increased counteracting the build-up of a population
inversion. The values of the PI in both schemes show a similar dependence
of the doping. A minor difference arises at $\sim\unit[100]{meV}$
and higher dopings, where the PI in scheme A does not decrease as
fast as the PI in scheme B. Here, the Fermi energy takes similar values
as the energy of $\text{LL}_{+2}$ and the direct pumping into $\text{LL}_{+2}$
in scheme A is beneficial, since Coulomb scattering from $\text{LL}_{+8}$
to $\text{LL}_{+2}$ in scheme B is strongly blocked in the beginning
of pumping due to $\rho_{+8}(t_{0})\ll\rho_{+2}(t_{0})$.

In general, comparing both PI schemes (cf. Fig. \ref{fig:gain-sketch}),
we find that they show a similar qualitative behavior. The PI rapidly
builds up already during the excitation and slowly decays after the
pump pulse on a time scale of few tens of ps (cf. Fig. \ref{fig:occupations}).
Furthermore, scheme B, describing a four-level laser system, is advantageous
for continuous laser action. This is based on the fact that a resonance
of an optical phonon energy with the transition $\text{LL}_{+1}\rightarrow\text{LL}_{-7}$
is feasible. This allows an electron to perform cycles in the four-level
system: First it is excited from $\text{LL}_{-7}$ to $\text{LL}_{+8}$,
from where it scatters down and accumulates in $\text{LL}_{+2}$,
before it participates in a stimulated emission event that transfers
it to $\text{LL}_{+1}$. Now, a phonon with the appropriate energy
can bring the electron back to $\text{LL}_{-7}$. The resonance condition
is fulfilled for the $\Gamma\text{TO}$, $\Gamma\text{LO}$, and $\text{KTO}$-phonon
modes at the magnetic fields $B=\unit[1.99]{T}$, $B=\unit[2.11]{T}$,
and $B=\unit[1.41]{T}$, respectively. A magnetic field of $B=\unit[2]{T}$
seems to be optimal, since no interfering resonances with other inter-LL
transitions occur (cf. Fig. \ref{fig:B-dependence}). 

Finally, we propose a simple pump-probe experiment to test
our predictions at sufficiently low temperatures, so that the initial
occupations $\rho_{+1}(t_{0})$ and $\rho_{+2}(t_{0})$ are nearly
zero: Exciting Landau-quantized graphene according to one of the PI
schemes A or B, cf. Fig. \ref{fig:gain-sketch}, with a probe pulse
measuring the $\sigma^{+}$- PI absorption, a positive differential
transmission signal (DTS) indicates a faster increase of $\rho_{+2}$
in comparison to $\rho_{+1}$ and consequently would provide strong
evidence for the occurrence of gain. 

In conclusion, based on microscopic calculations we predict the occurrence
of a pronounced population inversion in Landau-quantized graphene
and propose two different mechanisms to realize graphene-based LL
lasers. Surprisingly, we show that carrier-phonon scattering, which
generally reduces the population inversion, can be exploited to boost
the effect and to even open the way to continuous wave laser operation.
Furthermore, we demonstrate that the efficiency of the population
inversion can be tuned via experimentally accessible quantities, such
as the strength of the magnetic field and of the optical excitation.
In particular, controlling the doping of the sample allows us to also
tune the polarization of the emitted photons. Our microscopic insights
into the carrier dynamics in Landau-quantized graphene can guide future
experiments towards the design of graphene-based Landau level lasers
or THz emitters.

\begin{acknowledgments}
We acknowledge the financial support from the Einstein Stiftung Berlin
and we are thankful to the DFG for support through SPP 1459. Furthermore,
we thank A. Knorr (TU Berlin), and S. Winnerl (Helmholtz-Zentrum
Dresden-Rossendorf) for inspiring discussions on Landau-level lasers.
\end{acknowledgments}

\end{document}